\documentclass[aps,amssymb,prb,twocolumn,superscriptaddress]{revtex4}
\usepackage{graphicx}

\begin{document}

\title{Magnon Heat Conductivity and Mean Free Paths in Two-Leg Spin Ladders: A Model-Independent Determination}

\author{C. Hess}
\email[]{c.hess@ifw-dresden.de} \affiliation{IFW-Dresden, Institute for Solid State
Research, P.O. Box 270116, D-01171 Dresden, Germany}
\author{P. Ribeiro}
\affiliation{IFW-Dresden, Institute for Solid State Research, P.O. Box 270116, D-01171
Dresden, Germany}
\author{B. B\"uchner}
\affiliation{IFW-Dresden, Institute for Solid State Research, P.O. Box 270116, D-01171
Dresden, Germany}
\author{H. ElHaes}
\affiliation{2. Physikalisches Institut, RWTH-Aachen, 52056 Aachen, Germany}
\author{G. Roth}
\affiliation{Institut f\"{u}r Kristallographie, RWTH-Aachen, 52056 Aachen, Germany}
\author{U. Ammerahl}
\affiliation{Laboratoire de Physico-Chimie de l'\'{E}tat Solide, Universit\'{e} Paris-Sud, 91405 Orsay, France}
\author{A. Revcolevschi}
\affiliation{Laboratoire de Physico-Chimie de l'\'{E}tat Solide, Universit\'{e} Paris-Sud, 91405 Orsay, France}

\date{\today}

\begin{abstract}
The magnon thermal conductivity $\kappa_{\mathrm{mag}}$ of the spin ladders in $\rm
Sr_{14}Cu_{24-x}Zn_xO_{41}$ has been investigated at low doping levels $x=0$, 0.125, 0.25,
0.5 and 0.75. The Zn-impurities generate nonmagnetic defects which define an upper limit
for $l_{\mathrm{mag}}$ and therefore allow a clear-cut relation between $l_{\mathrm{mag}}$
and $\kappa_{\mathrm{mag}}$ to be established independently of any model. Over a large
temperature range we observe a progressive suppression of $\kappa_{\mathrm{mag}}$ with
increasing Zn-content and find in particular that with respect to pure $\rm
Sr_{14}Cu_{24}O_{41}$ $\kappa_{\mathrm{mag}}$ is strongly suppressed even in the case of
tiny impurity densities where $l_{\mathrm{mag}}\lesssim 374$~{\AA}. This shows
unambiguously that large $l_{\mathrm{mag}}\approx 3000$~{\AA} which have been reported for
$\rm Sr_{14}Cu_{24}O_{41}$
and $\rm La_{5}Ca_9Cu_{24}O_{41}$ on basis of a kinetic model are in the correct order of magnitude.%
%
\end{abstract}


\maketitle


\section{Introduction}

The transport of energy and magnetization in low dimensional quantum magnets is subject to numerous
experimental and theoretical
investigations,\cite{Sologubenko00,Hess01,Kudo01,Hess02,Hess04a,Sologubenko00a,Sologubenko01,Hess03,Hofmann03,Sologubenko03,Sologubenko03a,Zotos97,Alvarez02,Klumper02,Sakai03,Heidrich02,Heidrich03,Shimshoni03,Zotos04,Saito03,Orignac03,Heidrich04,Meier03,Heidrich05,Rozhkov05,Rozhkov05a}
with the nature of scattering and dissipation of thermal and magnetic currents at the center of
interest. In particular, quasi one-dimensional (1D) systems are being intensely studied since
properties starkly different from classical systems are expected: \textit{ballistic} magnetic heat
transport, i.e. a finite thermal Drude weight $D_\mathrm{th}$, is predicted in \textit{integrable}
models like the {\em XXZ} Heisenberg-chain with
$S=1/2$.\cite{Zotos97,Klumper02,Sakai03,Heidrich02,Heidrich03,Heidrich05} Experimentally, for the
$S=1/2$ two-leg spin ladder systems $\rm Sr_{14}Cu_{24}O_{41}$ and $\rm La_{5}Ca_9Cu_{24}O_{41}$
exceptionally large values of the magnetic thermal conductivity $\kappa_{\mathrm{mag}}$ are found
(up to $\sim 100~\rm Wm^{-1}K^{-1}$ even at room
temperature).\cite{Sologubenko00,Hess01,Kudo01,Hess02,Hess04a} In principle, these data should
provide the unique possibility of extracting the magnetic mean free path $l_{\mathrm{mag}}$, hence
obtaining information about magnetic scattering processes in such low-dimensional spin systems for
the first time. It turned out, however, that the extracted values of $l_{\mathrm{mag}}$ are
strongly model dependent: on one hand, the analysis within the framework of a simple kinetic model
yielded values up to $l_{\mathrm{mag}}\approx 3000$~{\AA}.\cite{Hess01} For the complicated and
supposedly rather imperfect materials studied such values are surprisingly large and might arise
from oversimplifications inherent to the model.
On the other hand, much smaller values were obtained in a more sophisticated analysis which
is based on a finite $D_\mathrm{th}$.\cite{Alvarez02} However, the finiteness of
$D_\mathrm{th}$ in these systems is subject of a highly controversial debate since both
ballistic and non-ballistic heat transport has been predicted for \textit{non-integrable}
$S=1/2$ models such as these two-leg spin-ladders or frustrated chains.\cite{Alvarez02,Heidrich02,Heidrich04,Heidrich03,Zotos04,Saito03,Orignac03,Shimshoni03}.

In this article we present an investigation of $\kappa_{\mathrm{mag}}$ and
$l_{\mathrm{mag}}$ of $\rm Sr_{14}Cu_{24}O_{41}$ which is independent of any model. The
basic idea is to generate nonmagnetic defects in the material and thereby impose an upper
limit for $l_{\mathrm{mag}}$ in a well-defined manner. We achieved this by substituting
small amounts of nonmagnetic Zn$^{2+}$ for the magnetic Cu$^{2+}$ (with $S=1/2$).
$\kappa_{\mathrm{mag}}$ of $\rm Sr_{14}Cu_{24-x}Zn_xO_{41}$ ($x\leq0.75$) is progressively
suppressed with increasing Zn-content and, in particular, already at tiny impurity
densities yielding $l_{\mathrm{mag}}\lesssim 374$~{\AA} this suppression is strong with
respect to the pure material (exceeding a factor of $\sim 5$). Employing the simple kinetic
model which had been used in previous studies\cite{Hess01,Hess04a} we even find
$l_{\mathrm{mag}}\approx d_\mathrm{Zn-Zn}$ at low temperatures, where $d_\mathrm{Zn-Zn}$ is
the nominal mean distance between Zn-ions along a ladder. This unambiguously shows that the
large $l_{\mathrm{mag}}\approx 3000$~{\AA} which have been reported for $\rm
Sr_{14}Cu_{24}O_{41}$ and $\rm La_{5}Ca_9Cu_{24}O_{41}$ are in the correct order of
magnitude.


\section{Experimental}
We have grown single crystals of $\rm Sr_{14}Cu_{24-x}Zn_xO_{41}$ ($x=0$, 0.125, 0.25, 0.5, 0.75)
by the travelling solvent floating zone method.\cite{Ammerahl98} The thermal conductivity as a
function of temperature $T$ of these crystals has been measured with a focus on the
crystallographic $c$-direction ($\kappa_c$), which is parallel to the orientation of the
ladders.\footnote{We note that at all doping levels we observed a tendency of $\kappa_c$ to vary
from measurement to measurement. Similar behavior has frequently been found in other quasi
one-dimensional spin systems and will be discussed elsewhere. Nevertheless, we stress that all data
presented here were reproducible in two subsequent measurements.} The measurements in the range
7-300~K have been performed with a standard four probe technique.\cite{Hess99,Hess01,Hess03a}

\begin{figure}
\includegraphics [width=1\columnwidth,clip] {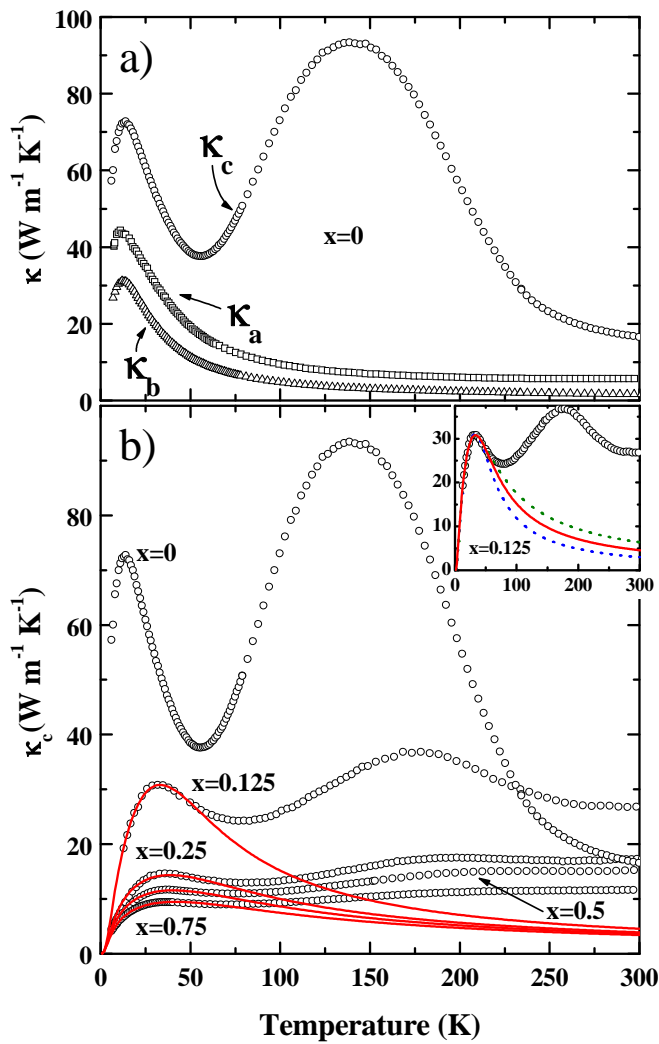}
\caption{\label{fig1}Thermal conductivity of $\rm Sr_{14}Cu_{24-x}Zn_xO_{41}$ ($x=0$,
0.125, 0.25, 0.5, 0.75) as a function of temperature. a) Anisotropic thermal conductivity
along $a$-, $b$-, and $c$-axes ($\kappa_a$, $\kappa_b$, $\kappa_c$) at $x=0$ (cf.
Ref.~\onlinecite{Hess01,Hess04a}). b) $\kappa_c$ at $x\geq0$. Solid lines represent the
estimated phononic background of $\kappa_c$ (cf. text). Inset: Estimations for the
uncertainties of the phonon background in the representative example $\kappa_c$ at
$x=0.125$ (dotted lines).}
\end{figure}

\section{Results and Analysis}
Fig.~\ref{fig1}~a) shows the temperature dependence of the thermal conductivity of the undoped
material, i.e. $\rm Sr_{14}Cu_{24-x}Zn_xO_{41}$ at $x=0$ along all three crystal axes. In this
material the thermal conductivity along the ladder direction, $\kappa_c(T)$, frequently displays a
double peak structure.\cite{Sologubenko00,Hess01,Kudo01,Hess02,Hess04a} The low-temperature peak
($\sim20$~K) is observed along all crystal directions and arises due to the usual phonon heat
conduction. The peak at higher temperature, however, is only observed along the $c$-axis and is the
signature of strictly one-dimensional magnetic heat conduction along the ladders in the
material.\cite{Sologubenko00,Hess01} In the following we concentrate on the evolution of $\kappa_c$
at finite Zn doping levels $x>0$ which is depicted in Fig.~\ref{fig1}~b). Upon Zn-doping,
$\kappa_c$ is progressively suppressed and exhibits less pronounced peaks. The magnetic peak
simultaneously broadens and apparently shifts towards higher temperature since the peak structure
evolves into a monotonic increase at high temperature.
%

Qualitatively, the doping induced suppression of both phononic and magnetic transport
channels of $\kappa_c$ is a natural consequence of the induced structural and magnetic
defects: different constituents on the same lattice site (Cu and Zn) must lead to enhanced
scattering of phonons. Similarly, since the Zn$^{2+}$ ions create nonmagnetic sites in a
$S=1/2$ spin ladder built of predominantly Cu$^{2+}$ ions, magnetic excitations scatter on
these sites. The upper limits for $l_{\mathrm{mag}}$ thus created correspond to the mean
distances between a pair of Zn-ions along a ladder, i.e. $d_\mathrm{Zn-Zn}\approx
374~${\AA}, 187~{\AA}, 94~{\AA} and 62~{\AA}. Apparently, even the small Zn-contents
discussed here lead to an efficient suppression of $\kappa_{\mathrm{mag}}$ with respect to
the undoped material. We can therefore conclude without further analysis that
$l_{\mathrm{mag}}$ of pure $\rm Sr_{14}Cu_{24}O_{41}$ must be significantly larger than the
highest $d_\mathrm{Zn-Zn}\approx 374~${\AA}.

Our aim is now to analyze the magnetic heat conductivity $\kappa_{\mathrm{mag}}$ of the ladders in
more detail; the phononic contribution $\kappa_{\mathrm{ph}}$ to $\kappa_c$ must therefore be
estimated and subtracted from $\kappa_c$. The estimation of $\kappa_{\mathrm{ph}}$ is based on the
fact that due to the large spin gap of order $\sim400$~K,\cite{Eccleston98,Katano99}
$\kappa_{\mathrm{mag}}$ is negligibly small at $T\lesssim 50$~K and hence
$\kappa_{\mathrm{ph}}\approx\kappa_c$ in this temperature range. Therefore, the usual
procedure\cite{Sologubenko00,Hess01,Hess04a} is to fit $\kappa_c$ at $T\lesssim 50$~K with a Debye
model\cite{Callaway60} and then extrapolate the fit function towards higher temperature. In our
analysis of $\kappa_{\mathrm{ph}}$ in the Zn-doped compounds we impose the additional constraint of
{\em identical fit parameters for all doping levels} provided that they are not related to
phonon-defect scattering. The coefficient for phonon-defect scattering is allowed to vary with
increasing Zn-content\footnote{For the fit we used\cite{Callaway60}
$\kappa_{\mathrm{ph}}=\frac{k_B}{2\pi^2v}\left(\frac{k_BT}{\hbar}\right)^3 \int_0^{\Theta_D/T}
\frac{\tau_cx^4e^x}{\left(e^x-1\right)^2}\,dx$ with the scattering rate
$\tau_c^{-1}=A\omega^{4}+v/L+BT\omega ^{3}\exp(-\Theta_D/bT)$, where $\omega$, $\Theta_D$ and $v$
are the phonon frequency, Debye temperature and mean phonon velocity. The three terms correspond to
scattering due to defects, crystal boundaries and umklapp processes respectively. $A$, $L$, $B$ and
$b$ are free fit parameters with the constraint that $L$, $B$ and $b$ are identical for all
Zn-doped samples.} and the results of this procedure are shown in Fig.~\ref{fig1}~b) as solid
lines. Such an extrapolation of $\kappa_{\mathrm{ph}}$ is certainly not exempt from uncertainty. We
take these into account by estimating positive and negative deviations from
$\kappa_{\mathrm{ph}}$.\footnote{The different phononic backgrounds have been obtained from fitting
data points at $T\leq (53\pm12)$~K.} A representative example of such deviation is shown in the
inset of Fig.~\ref{fig1} for $x=0.125$.


\begin{figure}
\includegraphics [width=1\columnwidth,clip] {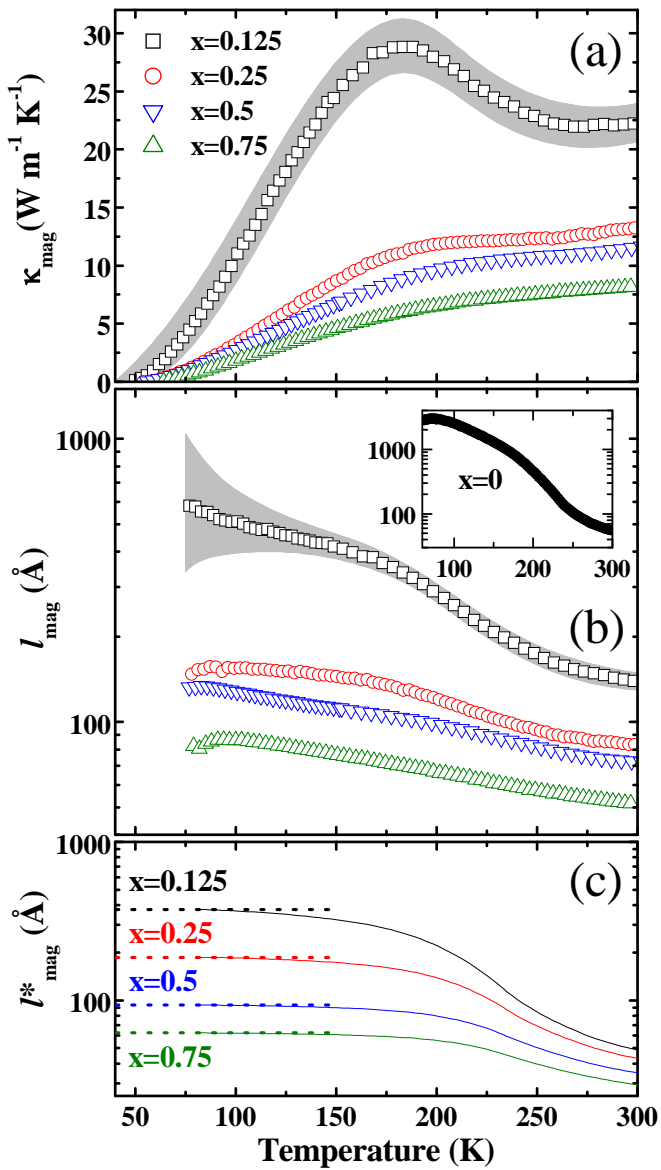}
\caption{\label{fig2}$\kappa_{\mathrm{mag}}$ (a) and $l_{\mathrm{mag}}$ (b) of $\rm
Sr_{14}Cu_{24-x}Zn_xO_{41}$ at $x=0.125$, 0.25, 0.5, 0.75 and $x=0$ (Inset). In (b) data points
below 75~K are omitted, because the experimental errors are to large in this temperature range. The
grey shaded areas in (a) and (b) display the experimental uncertainty of $\kappa_{\mathrm{mag}}$
and $l_{\mathrm{mag}}$ resulting from the uncertainties of $\kappa_{\mathrm{ph}}$ for the
representative case $x=0.125$. The solid lines in (c) represent the estimated $l^*_{\mathrm{mag}}$
(cf. text). The dotted lines mark the mean distance of Zn-ions in the ladders.}
\end{figure}

Fig.~\ref{fig2}a displays the resulting $\kappa_{\mathrm{mag}}$ of the four Zn-doped
compounds as a function of temperature. The peak-like shape apparent in the curve at
$x=0.125$ reflects the typical temperature dependence of the thermal conductivity of a
system with a moderate number of impurities: at low temperature, where scattering rates are
only weakly temperature dependent, the strong increase in the number of quasiparticles
generates an increasing $\kappa$. The decrease of $\kappa$ at intermediate temperature
($180~{\rm K}\lesssim T\lesssim 250$~K) and the almost constant to slightly increasing
$\kappa$ at further rising temperature signal a strong increase of the scattering rate and
a weakening or even saturation of this increase, respectively. The relevant scattering
processes for $\kappa_{\mathrm{mag}}$ in the undoped compound $\rm Sr_{14}Cu_{24}O_{41}$
are scattering on defects at low temperature and magnon-hole scattering at intermediate and
high temperatures.\cite{Hess02,Hess04a} In the case of our weakly doped samples with a
Zn-concentration $\lesssim 3\%$ it is reasonable to assume that this situation is in
principle preserved with the particularity, however, that the defects relevant for the
magnon scattering are mainly defined by the Zn-dopants. In this scenario one expects the
temperature independent magnon-defect scattering to increasingly dominate over the
temperature dependent magnon-hole scattering, when the Zn-doping, i.e. the density of
defects is further increased. $\kappa_{\mathrm{mag}}$ evolves accordingly upon doping,
since $\kappa_{\mathrm{mag}}$ decreases and the peak-like structure vanishes at
$x\gtrsim0.25$.

We proceed now with a quantitative analysis of $\kappa_{\mathrm{mag}}$, where our main
interest lies in a comparison of the mean distance of Zn-ions $d_\mathrm{Zn-Zn}$ and
$l_{\mathrm{mag}}$ extracted from the data. The analysis is based on a kinetic description
of the heat transport in the ladders presented earlier in Ref.~\onlinecite{Hess01},
\begin{equation}\label{fitmag}
\kappa_{\mathrm{mag}}=\frac{3 N l_{\mathrm{mag}}}{\pi\hbar k_BT^2}\int_{\Delta}^{\epsilon_{\mathrm{max}}}
\frac{\exp(\epsilon/k_BT)}{(\exp(\epsilon/k_BT)+3)^2}\epsilon^2d\epsilon .
\end{equation}
Here, $N$ is the number of ladders per unit area; $l_{\mathrm{mag}}$, $\Delta$ and
$\epsilon_{\mathrm{max}}\approx200$~meV\cite{Eccleston98} are the mean free path, the gap and the
band maximum of the spin excitations respectively. Presuming that the spin gap\cite{Hess01}
$\Delta=396$~K of undoped $\rm Sr_{14}Cu_{24}O_{41}$ does not significantly change at these light
Zn-doping levels,\footnote{Despite other Zn-doped low-dimensional quantum spins
systems\cite{Azuma97} our data on $\rm Sr_{14}Cu_{24-x}Zn_xO_{41}$ do not yield any evidence for
magnetic order at low temperature.} $l_{\mathrm{mag}}$ can directly be calculated from our
$\kappa_{\mathrm{mag}}$ data as shown in Fig.~\ref{fig2}b.

Similar to $l_{\mathrm{mag}}(T)$ at $x=0$ (cf. inset of Fig.~\ref{fig2}),
$l_{\mathrm{mag}}$ of the Zn-doped samples clearly decreases upon heating. The decrease
itself, however, is much weaker for the doped samples, and apparently weakens further with
increasing Zn-content. In Ref.~\onlinecite{Hess04a} it has been argued that in the undoped
compound the temperature dependence of $l_{\mathrm{mag}}$ is intimately related to
temperature dependent magnon-hole scattering which arises due to changes of hole mobility
upon charge ordering: at $T\lesssim100$~K, i.e. deep in the charge ordered state, magnon
hole scattering is unimportant. It becomes increasingly dominant at higher temperature and
reaches its full strength at $T\gtrsim240$~K, i.e. above the charge ordering temperature.
$l_{\mathrm{mag}}$ is thereby reduced from $l_0\approx3000$~{\AA} at $T\lesssim100$~K to
around 60~{\AA} at 300~K.

Following Matthiesen's rule it is possible to separate
$l_{\mathrm{mag}}^{-1}=l_0^{-1}+l_h^{-1}$ into contributions due to scattering off static
defects, $l_0$, and off mobile holes, $l_h$.\cite{Hess04a} The observed reduction of
$l_{\mathrm{mag}}$ upon Zn-doping is indeed consistent with a gradually reduced $l_0$ due
to scattering of magnons on Zn-ions and only a slight change in $l_h$. In order to
illustrate this, we estimate the mean free path of the Zn-doped samples under simplified
conditions, i.e. we assume that firstly $l_0=d_{\mathrm{Zn-Zn}}$ and secondly that $l_h(T)$
is the same as in the non-doped material as published in Ref.~\onlinecite{Hess04a}; hence
${l^*}_{\mathrm{mag}}^{-1}={d_{\mathrm{Zn-Zn}}}^{-1}+l_h^{-1}$.\footnote{Resistivity data
of $\rm Sr_{14}Cu_{24-x}Zn_xO_{41}$ indicate a weakening of the charge-ordered state upon
Zn-doping. An additional suppression of $l_{\mathrm{mag}}$ could therefore arise due to an
increased mobility of holes in the ladders and a thereby reduced $l_h$. However, this
effect appears to be only of minor importance, since a clear signature of charge ordering
is still observed at the lowest doping level $x=0.125$, whereas $l_{\mathrm{mag}}$ is
already strongly suppressed. In any case, the main doping- and temperature dependent
features of $l^*_{\mathrm{mag}}$ are quite robust against such doping dependencies as long
as $l_h\gg l_0$ at low temperature.} The results for $l^*_{\mathrm{mag}}$ are displayed in
Fig.~\ref{fig2}c. For $T\gtrsim100$~K there is a striking similarity in both magnitude and
temperature dependence between the results of the experimental $l_{\mathrm{mag}}$ and our
estimated $l^*_{\mathrm{mag}}$ values. We should mention the unavoidable uncertainties in
$l_{\mathrm{mag}}$ which are large at $T\lesssim100$~K, since the ratio
$\kappa_{\mathrm{mag}}/\kappa_{\mathrm{ph}}$ is small here and errors in the estimation of
$\kappa_{\mathrm{ph}}$ have large effects on the estimations of $\kappa_{\mathrm{mag}}$ and
$l_{\mathrm{mag}}$ (cf. Fig.~\ref{fig2}a and \ref{fig2}b). A reliable determination of the
experimental $l_{\mathrm{mag}}$ can therefore only be expected at higher temperature.

\begin{figure}
\includegraphics [width=\columnwidth,clip] {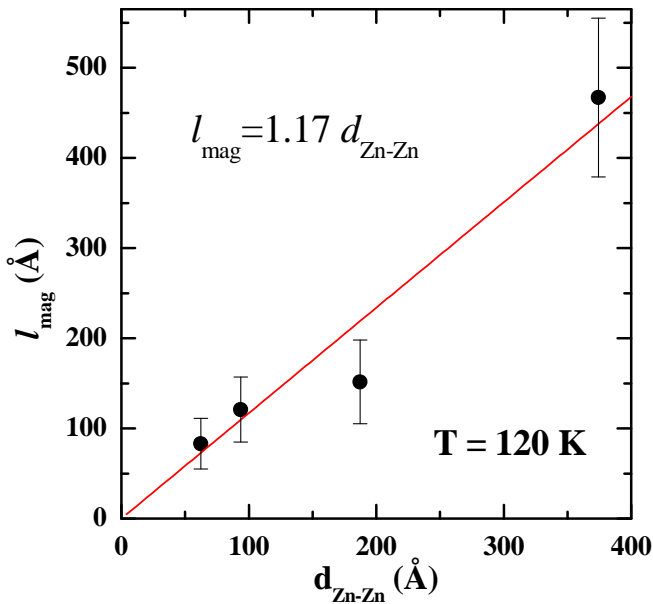}
\caption{\label{fig3}$l_{\mathrm{mag}}$ as a function of $d_{\mathrm{Zn-Zn}}$. Solid line: linear fit line through the
origin. The error bars arise due to uncertainties in determining $\kappa_{\mathrm {ph}}$ and have been obtained by
using different estimates for $\kappa_{\mathrm {ph}}$ (cf. inset of Fig.~\ref{fig1}) and Fig.~\ref{fig2}b.}
\end{figure}

However, at $100~{\rm K}\lesssim T\lesssim 125$~K such errors become reasonably small and
simultaneously $l_{\mathrm{mag}}\approx d_{\mathrm{Zn-Zn}}$ should hold. (Note that in this
temperature range the deviation of $l^*_{\mathrm{mag}}$ from $l_0=d_{\mathrm{Zn-Zn}}$ is smaller
than $10\%$, as can be inferred from the dotted lines in Fig.~\ref{fig2}c.) We therefore test our
experimental $l_{\mathrm{mag}}$ from Fig.~\ref{fig2}b for such a linear scaling at a representative
$T=120$~K in Fig.~\ref{fig3}.
As is evident from the figure, $l_{\mathrm{mag}}$ roughly scales with $d_{\mathrm{Zn-Zn}}$ and a linear fit to the data
points yields a slope of $1.17\pm 0.25$, which is close to unity. This confirms unambiguously that the values for
$l_{\mathrm{mag}}$ obtained from the kinetic model are in fair agreement with real scattering lengths in the material.

\section{Discussion}
The consequences of the analysis outlined above are evident. It is natural to conclude that the
kinetic model yields realistic numbers for $l_{\mathrm{mag}}$, not only in the Zn-doped material
discussed above but also in $\rm Sr_{14}Cu_{24}O_{41}$ and $\rm La_{5}Ca_9Cu_{24}O_{41}$ where it
yields the aforementioned surprising high values of $l_{\mathrm{mag}}\approx3000$~{\AA} at low
temperature. Therefore, the rather small values for $l_{\mathrm{mag}}$, which were obtained under
the assumption of a finite thermal Drude weight\cite{Alvarez02} become less realistic. This is
further reinforced by recent theoretical investigations which suggest a vanishing thermal Drude
weight\cite{Zotos04,Heidrich04}.

We point out that on a microscopic level the mechanism of magnon-scattering off nonmagnetic Zn$^{2+}$ ions and its
relevance for magnetic heat transport remains unaddressed. Qualitatively this scattering appears to be very similar to
the 2D case of $\rm La_2CuO_4$, where $l_{\mathrm{mag}}\approx d_{\mathrm{Zn-Zn}}$ has been observed as well (with
$d_{\mathrm{Zn-Zn}}$ being the mean unidirectional distance of Zn-impurities in the $\rm CuO_2$-planes).\cite{Hess03}
In this regard it seems to be highly intriguing to investigate the heat transport of Zn-doped spin chain materials such
as $\rm SrCuO_2$ and $\rm Sr_2CuO_3$, since Zn-impurities must lead to complete chain interruptions in these cases
which could -- compared to two-leg ladders and 2D antiferromagnetic planes -- result in a qualitatively different effect on $\kappa_{\mathrm{mag}}$.\\



\section{Summary}
Our experimental study of $\kappa_{\mathrm{mag}}$ of $\rm
Sr_{14}Cu_{24-x}Zn_xO_{41}$ shows that Zn-doping in this spin ladder compound strongly suppresses
$\kappa_{\mathrm{mag}}$ with a progressive decrease upon increasing Zn-content. Since the Zn-ions
create defects with a well defined distance the data provide a clear-cut and model-independent
relation between $\kappa_{\mathrm{mag}}$ and the mean free path $l_{\mathrm{mag}}\lesssim374$~{\AA}.
This strongly corroborates previous results for $l_{\mathrm{mag}}$ of pure $\rm
Sr_{14}Cu_{24}O_{41}$ and $\rm La_{5}Ca_9Cu_{24}O_{41}$, in which surprisingly large values for
$l_{\mathrm{mag}}$ have been found on basis of kinetic model\cite{Hess01}. We further find that the
$l_{\mathrm{mag}}$ determined from the present data with this kinetic model are roughly equal to
the mean distance of Zn-ions along the ladders. Magnetic scattering lengths in the spin ladder
material $\rm (Sr,Ca,La)_{14}Cu_{24}O_{41}$ can therefore conveniently be deduced from experimental
data for $\kappa_{\mathrm{mag}}$.

\begin{acknowledgments}
This work has been supported by the DFG through SP1073. It is a pleasure to thank W.
Brenig, F. Heidrich-Meisner, A. Rosch and A.V. Sologubenko for stimulating discussions. We
are in particular grateful to F. Hei\-drich-Meisner and R. Klingeler for carefully reading
the manuscript and useful comments. We further thank A.P. Petrovic and A. Waske for
proofreading the manuscript.
\end{acknowledgments}

\end{document}